\documentclass[aps,prd,noeprint,showpacs,twocolumn,amsmath,amssymb,superscriptaddress,nofootinbib]{revtex4-2}

\usepackage[varg]{txfonts}
\usepackage[]{graphicx}
\usepackage{enumerate}
\usepackage{subeqnarray}
\usepackage{cases}
\usepackage{mathrsfs}
\usepackage[usenames]{color}
\usepackage{bibentry}
\usepackage{hyperref}
\hypersetup{colorlinks=true, citecolor=blue, urlcolor=blue, linkcolor=blue}

\providecommand{\aap}{Astron. Astrophys.}
\providecommand{\mnras}{Mon. Not. R. Astron Soc.}
\providecommand{\apjl}{Astrophys. J. Lett}
\providecommand{\ssr}{Space Science Reviews}
\providecommand{\apss}{Astrophy. Space Sci.}
\providecommand{\jcap}{J. Cosmology Astropart. Phys.}
\providecommand{\apjs}{Astrophys. J. Suppl. Ser.}
\providecommand{\araa}{Annu. Rev. Astron. Astrophys.}
\providecommand{\physrep}{Phys. Rep.}

\begin{document} 
\author{D.~O.~Chernyshov}
\email{chernyshov@lpi.ru}
\affiliation{I.~E.~Tamm Theoretical Physics Division of P.~N.~Lebedev Institute of Physics, 119991 Moscow, Russia}
\author{A.~V.~Ivlev}
\email{ivlev@mpe.mpg.de}
\affiliation{Max-Planck-Institut f\"ur extraterrestrische Physik, 85748 Garching, Germany}
\author{A.~M.~Kiselev}
\affiliation{I.~E.~Tamm Theoretical Physics Division of P.~N.~Lebedev Institute of Physics, 119991 Moscow, Russia}

\title{Self-consistent theory of cosmic ray penetration into molecular clouds: relativistic case}

\nocite{skill76,cesarsky1978,everett11,morlino15,ivlev18, Phan2018, Bustard2021,Indriolo2012, indriolo15, Bacalla2019, Sabatini2023, Padovani2020, black1973, Issa1981, Aharonian1991, Tibaldo2021, Ackermann2012, Ackermann2013, yang2014, neronov17, remy2017,Ackermann201b, Calore2015, Gaggero2015, Yang2015, Acero2016, Ajello2016, Daylan2016,Huang2021,yang2023}

\date{\today}
\setcitestyle{authoryear,round}
\begin{abstract}
We study penetration of interstellar cosmic rays (CRs) into molecular clouds surrounded by nonuniform diffuse envelopes. The present work generalizes our earlier model of CR self-modulation \citep{ivlev18, dogiel2018}, in which the value for the envelope's gas density where CRs excite MHD waves was treated as a free parameter. Now, we investigate the case where the density monotonically increases toward the center. Assuming that CRs are relativistic, we obtain a universal analytical solution which does not depend on the particular shape of gas distribution in the envelope, and self-consistently derive boundaries of the diffusion zone formed within the envelope, where CRs are scattered at the self-excited waves. The values of the gas density at the boundaries are found to be substantially smaller than those assumed in the earlier model, which leads to a significantly stronger modulation of penetrating CRs. We compute the impact of CR self-modulation on the gamma-ray emission, and show that the results of our theoretical model are in excellent agreement with recent observations of nearby giant molecular clouds by \citet{yang2023}.
\end{abstract}
\setcitestyle{numbers,square}


   
   \maketitle
%

\section{Introduction}

Understanding penetration of interstellar cosmic rays (CRs) into molecular clouds is a long-standing problem in astrophysics. A general consensus is that MHD waves resonantly excited by streaming CRs in diffuse envelopes surrounding the clouds can inhibit their free penetration. As a result, the CR spectrum in the cloud interior may be reduced at energies contributing to the wave excitation (see, e.g., Refs.~\citep{skill76,cesarsky1978,everett11,morlino15,ivlev18, Phan2018, Bustard2021}). For a long time, this phenomenon of CR self-modulation has been of merely theoretical interest, but recent rapid development in observational astronomy has made it possible to start testing the theory against the acquired data. 

Numerous data available on the CR ionization rate in molecular clouds, derived from observed abundances of various ions generated due to CR ionizing collisions with gas, clearly shows a tendency for the ionization rate to decrease with the cloud column density (see Refs.~\citep{Indriolo2012, indriolo15, Bacalla2019, Sabatini2023} and references therein). While such behavior generally agrees with theoretical expectations of CR attenuation due to ionization losses (see, e.g., Ref.~\citep{Padovani2020}), it may also reflect the effect of CR self-modulation that should be most pronounced at lower energies. However, it is rather difficult to discriminate between the two mechanisms using these data -- on the one hand, the attenuation models contain many poorly constrained parameters; on the other hand, the ionization rate alone cannot be used to identify the predicted changes in the CR spectrum.

At the same time, gamma-ray diffuse emission produced by relativistic CRs interacting with dense gas has an ability to reveal the CR spectrum (see Refs.~\citep{black1973,Issa1981, Aharonian1991} and review by \citet{Tibaldo2021}). Measurements of gamma-ray spectra in individual molecular clouds is a challenging task, both for local clouds (see, e.g., Refs.~\citep{Ackermann2012, Ackermann2013, yang2014, neronov17, remy2017}) and for clouds in the Galactic center (see, e.g., Refs.~\citep{Ackermann201b, Calore2015, Gaggero2015, Yang2015, Acero2016, Ajello2016, Daylan2016}). Nevertheless, there are hints that the spectrum of relativistic CRs in the central molecular zone is suppressed compared to the surrounding ``sea'' spectrum \citep{Huang2021}. Furthermore, a recent analysis of the gamma-ray emission in the direction of nearby giant molecular clouds provides clear evidence that the emission from the dense clumps is reduced at GeV energies \citep{yang2023}. One of the possible mechanisms that can explain these observations is CR self-modulation.

Earlier, we have studied the effects of CR self-modulation on the gamma-ray emission from dense molecular clouds \citep{dogiel2018}. Our conclusion, consistent with the results by \citet{skill76}, was that the expected impact of self-modulation on the emission is typically marginal, and may be reliably detected only in very massive clouds with a column density well over $10^{23}$~cm$^{-2}$. However, in that paper we assumed the diffuse envelopes around molecular clouds to have a constant gas density, and treated its value as a free model parameter.

In the present paper, we generalize our earlier models \citep{ivlev18, dogiel2018} and study penetration of relativistic interstellar CRs into molecular clouds surrounded by nonuniform envelopes. The paper is organized as follows: in Sec.~\ref{sec:setup} we discuss the qualitative picture of the process and summarize the basic phenomenology, and also present the governing equations and the boundary conditions. A self-consistent solution of the problem for CR protons is given in Sec.~\ref{sec:sol_nolos}. In Sec.~\ref{sec:sol_multiple} we expand the solution for multiple CR nuclei, and in Sec.~\ref{sec:nucl_contribution} discuss the relative impact of individual nuclei. In Sec.~\ref{sec:gammarays} we compute the effect of CR self-modulation on the gamma-ray emission and compare the theoretical results with observational data by \citet{yang2023}. In Sec.~\ref{conclusions} we summarize the principal finding of the paper, and highlight the essential role of gas inhomogeneity in the self-modulation process. 

\section{Problem setup}\label{sec:setup}

In Ref.~\citep{ivlev18} we have shown that CRs penetrating dense molecular clouds excite MHD turbulence in diffuse envelopes of the clouds. This leads to the formation of a turbulent {\it diffusion zone}, where CRs are efficiently scattered at the self-excited waves. The diffusion zone has a crescent shape in the plane spanned by the CR energy/momentum and the depth, with the tip set by the {\it excitation threshold} -- the maximum energy above which the CR flux is insufficient to generate turbulence. Depending on the interstellar (ISM) spectrum and the cloud column density, the density of CRs in the cloud interior may be significantly modulated at energies below the excitation threshold \citep{ivlev18, dogiel2018}. However, in these papers we considered a simplified model of diffuse envelopes, assuming the gas density has a certain constant value $\sim10$~cm$^{-3}$ across the whole envelope. 

Recently developed dust extinction maps \citep{Leike2020, Edenhofer2024} allow us to reconstruct the 3D dust/gas distribution within $\approx1$~kpc proximity to the Sun, with a spatial resolution of 1~pc. These maps are therefore an excellent tool to resolve the gas distribution in envelopes of nearby molecular clouds: they show that gas is highly nonuniform, with the density typically varying from well below $\sim1$~cm$^{-3}$ in most of the ISM volume up to dozens of cm$^{-3}$ near the peaks of gas clumps (where unresolved sub-pc dense cores tend to concentrate). The maps suggest that gas inhomogeneities in the envelopes may play an essential role, and that typical density values relevant for wave excitation may be substantially lower than those assumed in Refs.~\citep{ivlev18} and \citep{dogiel2018}.

In the present paper we generalize the problem studied in \citet{ivlev18} and \citet{dogiel2018} to nonuniform envelopes. As demonstrated in the following sections, this step not only allows us to solve the problem self-consistently, without assuming the value for the gas density where CRs excite waves, but also makes it possible to obtain a general analytical solution for the diffusion zone and to derive the corresponding diffusion coefficient of relativistic CRs. 

Interstellar CRs penetrate molecular clouds along the local magnetic field lines, making the problem essentially one-dimensional. We assume the gas density $n$ in envelopes to increase monotonically toward dense clumps, versus the distance $z$ measured along the field lines. Thus, $n$ can be employed as a new coordinate instead of $z$: in Sec.~\ref{sec:sol_nolos} we show that the resulting solution becomes universal and independent on a particular shape of $n(z)$. 

We point out that outer regions of diffuse envelopes have negligible contribution to the total column density of molecular clouds. This implies that attenuation of relativistic CRs penetrating a cloud and, therefore, the net velocity of their flux into the cloud is completely determined by the column density of the dense clump, as given by Eq.~(\ref{eq:uN_definiton}) in Sec.~\ref{sec:gammarays}. At the same time, generation of self-excited waves by the net CR flux is only possible in diffuse envelopes, as the waves are efficiently damped at higher densities. Thus, the self-consistent treatment of the problem, including the solution for the diffusion zone, is reduced to analysis of the processes occurring in the envelope, while the dense clump only sets the value of the net flux velocity.

\begin{figure}[ht]\centering
\includegraphics[width=0.45\textwidth]{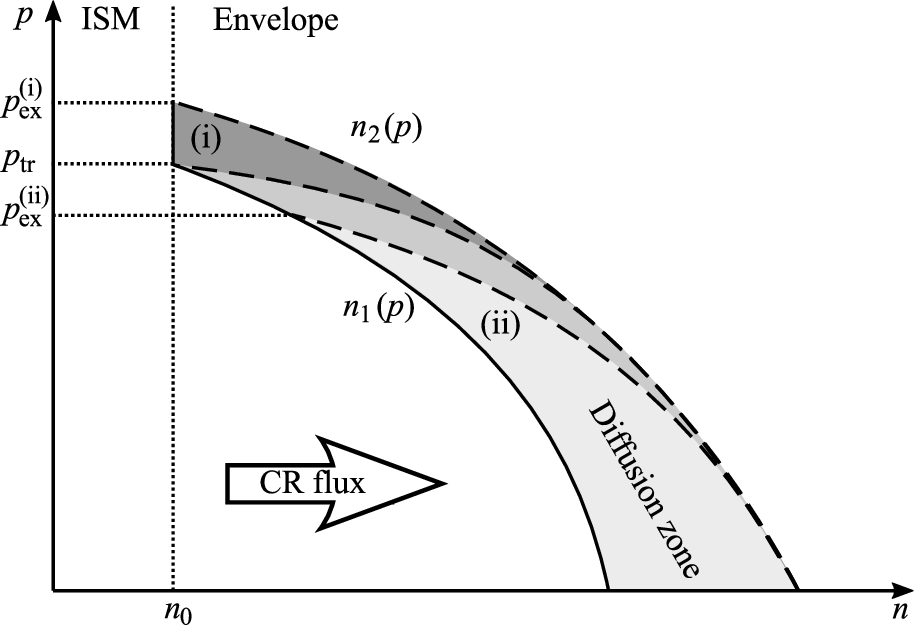}
\caption{A sketch illustrating penetration of CRs from the ISM into a nonuniform envelope of a molecular cloud. The border between the ISM and the envelope is set at the gas density $n=n_0$, CRs are able to excite Alfven waves at $n\geq n_0$. For a given CR momentum $p$, the region of self-excited turbulence is bound between the {\it outer} boundary $n_1(p)$ and the {\it inner} boundary $n_2(p)$ (depicted by the solid and dashed lines, respectively). These boundaries identify the {\it diffusion zone} for CRs in the $(p,n)$ plane (shaded region). Two regimes of the diffusion zone can be realized: (i) either both $n_1(p)$ and $n_2(p)$ end at $n=n_0$, or (ii) they intersect at $n > n_0$. The respective {\it excitation threshold} (i.e., the maximum momentum at which CRs are able to excite waves) is denoted by $p_{\rm ex}^{\rm (i)}$ and $p_{\rm ex}^{\rm (ii)}$, the transition between the two regimes is at $p_{\rm ex}=p_{\rm tr}$.} \label{fig:diagram}
\end{figure}

The diffusion zone in the $(p,n)$ plane is sketched in Fig.~\ref{fig:diagram}. While it appears qualitatively similar to that plotted in Fig.~2 of \citet{ivlev18}, the location of the zone boundaries as well as the underlying formation mechanisms turn out to be quite different. In Ref.~\citep{ivlev18}, the diffusion zone emerges near an artificial sharp border between a low-density envelope and a high-density clump. In the present model, the diffusion zone forms naturally, as a result of intrinsic inhomogeneity which controls the locations of the boundaries for a given $p$. 

We truncate a nonuniform diffuse envelope at a certain density $n_0$, which identifies a border through which CRs enter the envelope from the ISM. The border is located at a local minimum of the wave damping rate, which is associated with a transition between different dominant ions in the envelope and the ISM (see Sec.~\ref{governing}), such that penetrating CRs excite waves only at $n\geq n_0$. 

Fig.~\ref{fig:diagram} illustrates characteristic realizations of the diffusion zone for a given spectrum of interstellar CRs, plotted for three different values of the cloud column density $\mathcal{N}$. For a given $p$, the diffusion region is bound between the {\it outer} boundary $n_1(p)$ (solid line) and the {\it inner} boundary $n_2(p)$ (dashed lines). The position of $n_1$ depends on the magnetic field, the border gas density, the ionization fraction and composition of the gas, as well as the ISM spectrum. The column density -- which is a measure of CR attenuation in the cloud interior -- controls the relative position of $n_2$ with respect to $n_1$ and, hence, the value of the excitation threshold $p_{\rm ex}$. The latter is an increasing function of $\mathcal{N}$ \citep{ivlev18, dogiel2018}. Therefore, regime (ii) depicted in Fig.~\ref{fig:diagram} corresponds to a lower value of $\mathcal{N}$, regime (i) to a higher one, and the transition between the regimes to a medium one. As discussed in Sec.~\ref{sec:gammarays}, regime (i) is expected to be realized for nearby giant molecular clouds and the local spectrum of Galactic CRs.

\subsection{General governing equations}
\label{governing}

Propagation of CRs within the diffusion zone should take into account their interaction with the self-generated turbulence. This process is generally described by a set of coupled equations for the CR spectrum and the turbulent spectrum. The problem is similar to that studied in Refs.~\citep{ivlev18, dogiel2018}, but now gas inhomogeneity in the envelope is also included. 

Propagation of protons is governed by the advection-diffusion transport equation
(see, e.g., Refs.~\citep{skill76,ptus06, morlino15}):
\begin{equation}
\frac{\partial}{\partial z}\left( v_{\rm A}f - D\frac{\partial f}{\partial z}\right) - \frac{\partial}{\partial p}\left(\dot{p}f\right) = 0\,,
\label{eq:main_propagation}
\end{equation}
where $f(p,z)$ is the CR distribution function (spectrum) in the momentum space (normalized such that the local number density of protons is $\int f(p,z) dp$), and $\dot{p}>0$ describes continuous energy losses due to interaction with gas. The CR advection is set by the velocity of self-excited waves, equal to the Alfven velocity,
\begin{equation}
v_{\rm A}(z) = \frac{B}{\sqrt{4\pi m_i \xi_i n}}\,,
\end{equation}
which is determined by the longitudinal magnetic field $B(z)$, the gas density $n(z)$, and the mass $m_i$ of the dominant ion with the abundance $\xi_i$. We assume ionized carbon with $\xi_i = 1.5\times 10^{-4}$ \citep{DraineBook2011} to dominate in the envelope. The CR diffusion coefficient \citep{Skilling1975},
\begin{equation}
D(p,z) \simeq \frac{1}{6\pi^2}\frac{vB^2}{k^2W}\,,
\label{D}
\end{equation}
is determined by the spectral energy density of self-excited turbulence, $W(k,z)$, where the wavenumber $k$ is related to $p$ via the resonance condition $k = eB/pc$.

Following Refs.~\citep{skill76, ivlev18}, the wave equation for $W(k,z)$ in diffuse envelopes of molecular clouds can be reduced to the excitation-damping balance,
\begin{equation}
\gamma_{\rm CR}=\nu_{in} \,.
\label{Wave_Eq}
\end{equation}
The wave excitation rate by CRs, $\gamma_{\rm CR}$, is described by the following equation \citep{Skilling1975, morlino15}:
\begin{equation}
\gamma_{\rm CR}(k,z) \simeq - \pi^2 \frac{e^2v_{\rm A}}{m_pc^2\Omega_p} pD \frac{\partial f}{\partial z}\,,
\label{eq:value_of_gamma}
\end{equation}
where $\Omega_p = eB/m_p c$ is the non-relativistic girofrequency of protons. The rate of wave damping, $\nu_{in}$, is equal to one half of the momentum transfer rate for ion-neutral collisions,
\begin{equation}
\nu_{in}(z) = \frac12\frac{m_nn}{m_n+m_i}\langle\sigma_{in} v\rangle\equiv \nu_0\left(\frac{n}{n_0}\right) \,,
\label{eq:nudamp_linear}
\end{equation}
where $\langle\sigma_{in} v\rangle$ is the corresponding momentum-transfer rate coefficient. 
For atomic neutral hydrogen and carbon ions, we have $\langle\sigma_{in} v\rangle\approx 2.4\times10^{-9}$~cm$^3$/s \citep{DraineBook2011}.

In the outer envelope regions, where the gas density drops below $\sim1$~cm$^{-3}$, the dominant ions are H$^+$ with $\langle\sigma_{in} v\rangle\approx 3.3\times10^{-9}$~cm$^3$/s \citep{DraineBook2011}. For a given gas density, this change in the ion composition leads to an increase in the damping rate $\nu_{in}$ by a factor of $\approx8.9$. Therefore, we assume that turbulence exists only where carbon ions dominate, and the density $n=n_0$ where this change occurs identifies the border between the ISM and the diffuse envelope. For $n_0 = 1$ cm$^{-3}$, we obtain $\nu_0 \approx 9\times 10^{-11}~\mbox{s}^{-1}$.

The excitation-damping balance, Eq.~(\ref{Wave_Eq}), allows us to obtain the exact expression for the diffusion flux:
\begin{equation}
-D\frac{\partial f}{\partial z} = \frac{Bc\nu_{in}}{\pi^2ev_{\rm A}p} \equiv S_{\rm D}(p,z) \,.
\label{eq:sdd_definition}
\end{equation}
One can see that $S_{\rm D} \propto n^{3/2}$ and does not depend on $B$ explicitly. Eq.~(\ref{eq:sdd_definition}) represents the universal, {\it diffusion-dominated} flux of penetrating CRs. It was derived in \citet{ivlev18} for the limit where the advection contribution is small, and should not be confused with the {\it total} CR flux $S$ within the diffusion zone,  
\begin{equation}
S(p,z) = S_{\rm D} + v_{\rm A}f \,,
\label{eq:totals_definition}
\end{equation}
which is obtained by solving the transport equation (\ref{eq:main_propagation}). 

\subsection{Boundary conditions for relativistic CRs}\label{sec:boundaries}

Continuous losses $\dot{p}$ are proportional to the gas density and therefore play a minor role in diffuse envelopes. Hence, this term can be safely neglected at relativistic energies, which reduces Eq.~(\ref{eq:main_propagation}) to the total flux conservation,
\begin{equation}
S(p,z) = S_0(p) \,.
\end{equation}

To solve the problem, we need to determine the outer and inner boundaries of the diffusion zone, $z_1$ and $z_2$, and to set the boundary conditions. We assume no CR scattering outside the diffusion zone, and hence no CR gradient. Therefore, the outer boundary condition is
\begin{equation}
\left.f(p,z)\right|_{z = z_1} = f_0(p) \,,     
\end{equation}
where $f_0(p)$ is the ISM spectrum of CRs. For the inner boundary, we follow Ref.~\citep{dogiel2018} and introduce the net flux velocity $u(p)$, which is determined by the absorption of CRs in the interior of a molecular cloud. The boundary condition at $z=z_2$ is 
\begin{equation}
u(p)\left.f(p,z)\right|_{z = z_2} = S_0(p) \,.
\label{eq:z2_boundary}
\end{equation}
The value of $u$ increases with the total column density of the cloud (see Sec.~\ref{sec:gammarays}). 

For brevity, below we denote the values of parameters at $z = z_i$ by the corresponding indexes, e.g., $\left.n\right|_{z = z_1} = n_1$, $\left.v_{\rm A}\right|_{z = z_2} = v_{\rm A2}, \left.S_{\rm D}\right|_{z = z_2} = S_{\rm D2}$, etc. 

\section{Solution for protons}\label{sec:sol_nolos}

We assume $B=$~const, which is a very reasonable assumption in diffuse envelopes of molecular clouds \citep{Crutcher2012}. For a monotonically increasing $n(z)$, we write at $n>n_0$
\begin{equation}
f\frac{dv_{\rm A}}{dz} = -\frac{v_{\rm A}f}{2n} \frac{dn}{dz} = \frac{S_{\rm D} - S_0}{2n} \frac{dn}{dz} \,.
\label{dv_A/dz}
\end{equation}
Differentiating Eq.~(\ref{eq:totals_definition}) over $z$ and using Eq.~(\ref{dv_A/dz}), we arrive to
\begin{equation}
\frac{df}{dz} = \frac{S_0 - 4S_{\rm D}}{2v_{\rm A}n} \frac{dn}{dz} \,.
\label{eq:dfdz_definition}
\end{equation}

The diffusion coefficient must be finite within the diffusion zone, which requires $\gamma_{\rm CR} > 0$. According to Eq.~(\ref{eq:value_of_gamma}), this in turn requires $\partial f/\partial z < 0$. Hence, the outer boundary of the diffusion zone for $n(z)>n_0$, denoted by $z_{\rm cr}(p)$, is determined from the condition $\left.\partial f/\partial z\right|_{z = z_{\rm cr}} = 0$. Combining this with Eq.~(\ref{eq:dfdz_definition}), we obtain $S_0 = 4\left.S_{\rm D}\right|_{z = z_{\rm cr}} $, and by virtue of Eq.~(\ref{eq:totals_definition}) reduce it to 
\begin{equation}
3\left.S_{\rm D}\right|_{z = z_{\rm cr}} = \left.v_{\rm A}\right|_{z = z_{\rm cr}}f_0\,.
\label{eq:z1_boundary}
\end{equation}
Substituting $S_{\rm D}$ from Eq.~(\ref{eq:sdd_definition}), we finally derive the critical density $n_{\rm cr}(p)\equiv n(z)|_{z=z_{\rm cr}(p)}\,$,
\begin{equation}
n_{\rm cr}(p) \equiv \sqrt{\frac{\pi pf_0(p)n_0}{12\xi_i} \frac{\Omega_i}{\nu_0}}\,,
\label{eq:n1}
\end{equation}
where $\Omega_i = eB/m_ic$ is the ion gyrofrequency. If $n_{\rm cr}$ is smaller than $n_0$, we set $n_1 = n_0$. Thus, the outer boundary of the diffusion zone  is
\begin{equation}
n_1(p) = n_0 \max \left\{\frac{n_{\rm cr}(p)}{n_0}, 1\right\}\,.
\end{equation}

We write the total flux as
\begin{equation}
    S_0 = S_{\rm D1} + v_{\rm A1}f_0 \,,
\end{equation}
where
\begin{equation}
    v_{\rm A1}f_0 = \left.v_{\rm A}\right|_{z = z_{\rm cr}}f_0\left(\frac{n_{\rm cr}}{n_1}\right)^{1/2} = 3\left.S_{\rm D}\right|_{z = z_{\rm cr}}\left(\frac{n_{\rm cr}}{n_1}\right)^{1/2} \,.
\end{equation}
According to Eq.~(\ref{eq:sdd_definition}), $\left.S_{\rm D}\right|_{z = z_{\rm cr}} = S_{\rm D1}(n_{\rm cr}/n_1)^{3/2}$. Then, by introducing the following factor: 
\begin{equation}
    \mathcal{K}(p) = 3\left(\frac{n_{\rm cr}(p)}{n_1(p)}\right)^2 + 1\,,
    \label{K}
\end{equation}
we can express $S_0$ as
\begin{eqnarray}
    S_0(p) = \mathcal{K}S_{\rm D1} \label{eq:Snoloss_solution}\hspace{6.cm}\\
    \hspace{0cm} \left\{
    \begin{array}{ll}
         =4S_{\rm D}(n_{\rm cr})\propto p^{-1/4}f_0^{3/4}(p) &\mbox{if}~n_{\rm cr}(p) > n_0\,; \\
         \approx S_{\rm D}(n_0)\propto p^{-1} & \mbox{if}~n_{\rm cr}(p) \ll n_0/\sqrt{3}\,.
    \end{array}
    \right.
    \nonumber
\end{eqnarray}
Unlike the case of uniform envelopes studied in \citet{ivlev18}, where the diffusion term was shown to dominate the modulated flux at higher energies and the avection term -- at lower energies, now the magnitude of $S_0(p)$ is always set by the universal diffusion flux $S_{\rm D}$, determined by Eq.~(\ref{eq:sdd_definition}). 
                                                                                      
Using Eqs.~(\ref{eq:totals_definition}) and (\ref{eq:z2_boundary}) we rewrite the condition on the inner boundary of the diffusion zone as
\begin{equation}
    S_{\rm D2} = S_0 - v_{\rm A2}f_2 = S_0\left(1 - \frac{v_{\rm A2}}{u}\right) \,.
\end{equation}
Substituting $S_{\rm D2} = S_{\rm D1}(n_2/n_1)^{3/2}$ and using Eq.~(\ref{eq:Snoloss_solution}), we obtain an algebraic equation for $n_2$,
\begin{equation}
n_2(p) = n_1(p)\left[\mathcal{K}\left(1 - \frac{v_{\rm A2}}{u}\right)\right]^{2/3} \,.
\label{eq:n2}
\end{equation}
Solution of this equation is only valid for $n_2(p) > n_1(p)$, and hence the excitation threshold is given by the condition 
\begin{equation}
p_{\rm ex}:\qquad n_2(p_{\rm ex}) = n_1(p_{\rm ex})\,.
\end{equation}

Thus, the diffusion zone in the $(n,p)$ plane is completely characterized by two similarity numbers, $n_{\rm cr}(p)/n_0$ and $u(p)/v_{\rm A0}$. We note that $u$ at (sub)relativistic energies strongly depends on $p$ due to ionization losses, while for ultra-relativistic energies $u \approx$~const \citep{dogiel2018, dogiel2021}.

As illustrated in Fig.~\ref{fig:diagram}, two regimes of the diffusion zone can be realized: (i) either both $n_1(p)$ and $n_2(p)$ end at $n=n_0$, or (ii) they intersect at $n > n_0$. A transition between the regimes occurs if the two boundaries intersect exactly at $n = n_0$, i.e., when $n_1(p_{\rm ex}) = n_2(p_{\rm ex}) = n_0$. In this case $\mathcal{K}(p_{\rm ex}) = 4$, and from Eq.~(\ref{eq:n2}) we derive the following transition condition:
\begin{equation}
    v_{\rm A0} = \frac{3}{4}u(p_{\rm tr}) \,,
    \label{eq:n_trans}
\end{equation}
where the transition momentum (i.e., the value of $p_{\rm ex}$ at the transition point) is given by the condition 
\begin{equation}
p_{\rm tr}:\qquad n_{\rm cr}(p_{\rm tr})=n_0\,.  
\label{p_tr}
\end{equation}

Regime (i) is realized if $v_{\rm A0} < \frac{3}{4}u$. It is worth noting that $u$ is proportional to the total gas column density (see Sec.~\ref{sec:gammarays}), and therefore this regime is expected to operate in denser molecular clouds. The corresponding value of $p_{\rm ex}~(>p_{\rm tr})$, denoted by $p_{\rm ex}^{\rm (i)}$ in Fig.~\ref{fig:diagram}, is set by condition $n_2(p_{\rm ex})=n_0$. Eqs.~(\ref{K}) and (\ref{eq:n2}) yield
\begin{equation}
    \frac{n_{\rm cr}(p_{\rm ex})}{n_0} = \sqrt{\frac{v_{\rm A0}}{3[u(p_{\rm ex}) - v_{\rm A0}]}} \,.
    \label{p_ex1}
\end{equation}
Regime (ii) corresponds to $v_{\rm A0} > \frac{3}{4}u$. Then $p_{\rm ex}~(<p_{\rm tr})$, denoted by $p_{\rm ex}^{\rm (ii)}$ in Fig.~\ref{fig:diagram}, is derived from $n_1(p_{\rm ex})=n_2(p_{\rm ex})$. In this case, $\mathcal{K}(p_{\rm ex}) = 4$ and
\begin{equation}
    \left.v_{\rm A}\right|_{n = n_{\rm cr}(p_{\rm ex})} = \frac{3}{4}u(p_{\rm ex}) \,.
    \label{eq:pex_2variant}
\end{equation}
According to Eq.~(\ref{eq:n2}), the gas density variation within the diffusion zone always remains moderate: since $\mathcal{K} \leq 4$, we have $n_2/n_1 < \mathcal{K}^{2/3} \leq 2^{4/3} \approx 2.5$.

Finally, using Eqs.~(\ref{eq:sdd_definition}) and (\ref{eq:dfdz_definition}) we express $D$ within the diffusion zone $n_1 \leq n \leq n_2$ in the following form:
\begin{equation}
\frac{v_{\rm A}}{D} = \frac{1}{2n} \left[4 - \mathcal{K}\left(\frac{n}{n_1}\right)^{-3/2}\right]\frac{dn}{dz} \,,
\label{eq:D_noloss}
\end{equation}
The diffusion coefficient diverges at $n \to n_1$ if $n_1 = n_{\rm cr}$, but remains finite otherwise. The mean free path for protons, $\sim 3D/c$, is much smaller than the gas inhomogeneity scale $n/(dn/dz)\sim 2D/v_{\rm A}$.
Thus, the diffusion equation is always applicable within the diffusion zone.

From Eqs.~(\ref{eq:z2_boundary}) and (\ref{eq:Snoloss_solution}) it is evident that the total CR flux $S_0(p)$ and, hence, spectrum $f_2(p)$ in the cloud interior does not depend on the shape of $n(z)$, provided it is a monotonically increasing function. The results only depend on the border density $n_0$ if regime (i) is realized for the diffusion zone.

In Appendix~\ref{app2} we show that the obtained stationary solution is stable against small-scale perturbations. Furthermore, it is interesting to note that the value of total CR flux is minimized at the outer boundary of the diffusion zone: by differentiating Eq.~(\ref{eq:totals_definition}) over $z$ and keeping in mind that $S_{\rm D}\propto n^{3/2}$ and $v_{\rm A}\propto n^{-1/2}$, we immediately conclude that the total flux reaches the local minimum $S_0(p)$ at $z_{\rm cr}(p)$, as defined by Eq.~(\ref{eq:z1_boundary}) (or at $z_0$ if $z_{\rm cr}<z_0$). Thus, the position of the outer boundary is stable, too. Regarding stability of the inner boundary, we keep in mind that $f(p,z)$ decreases monotonically within the diffusion zone. Therefore, small positive (negative) variations in the boundary position relative to the equilibrium increase (decrease) the local CR gradient, and thus induce wave damping (excitation) pushing the boundary back to equilibrium.

\section{Solution for multiple nuclei}\label{sec:sol_multiple}

If elements heavier than protons are considered, it is more convenient to express spectra for individual nuclei, $f^{(\alpha)}$, in terms of the magnetic rigidity $R = pc/eZ^{(\alpha)}$. In the absence of losses, Eq.~(\ref{eq:main_propagation}) is then transformed into the following equations:
\begin{equation}
v_{\rm A}f^{(\alpha)} - D^{(\alpha)}\frac{\partial f^{(\alpha)}}{\partial z}= S_0^{(\alpha)}(R)\,,
\label{eq:main_propagation_nucl}
\end{equation}
where $Z^{(\alpha)}$ is the atomic number of species $\alpha$. The rigidity is convenient because the resulting resonance condition does not depend on the charge, $k = B/R$, and then the diffusion coefficient of relativistic CRs is a function of $R$ only, i.e., $D^{(\alpha)}(R) \equiv D(R)$ for all nuclei. The spectra are normalized in the way that $\int f^{(\alpha)}(R)~ dR$ is the number density of nuclei $\alpha$. 

Eq.~(\ref{eq:value_of_gamma}) for multiple CR species is rewritten as (see, e.g., Ref.~\citep{kuls69})
\begin{equation}
\gamma_{\rm CR}(k,z) \simeq - \pi^2 \frac{e^2v_{\rm A}}{m_pc^2\Omega_p} R\sum\limits_{\alpha} Z^{(\alpha)}D^{(\alpha)} \frac{\partial f^{(\alpha)}}{\partial z}\,.
\label{eq:value_of_gamma_nucl}
\end{equation}
Assume that all particle are relativistic. By multiplying the individual fluxes in Eq.~(\ref{eq:main_propagation_nucl}) by $Z^{(\alpha)}$ and summing them up, we arrive to a single-species problem with the effective ISM spectrum given by
\begin{equation}
f^{\Sigma}(R,z) = \sum\limits_{\alpha}Z^{(\alpha)}f^{(\alpha)}(R,z) \,. 
\end{equation}
The resulting equations to solve are
\begin{equation}
 v_{\rm A}f^{\Sigma} - D\frac{\partial f^{\Sigma}}{\partial z} = \sum\limits_{\alpha} Z^{(\alpha)}S_0^{(\alpha)}(R) \equiv S_0^\Sigma(R) \,,
\end{equation}
and
\begin{equation}
\gamma_{\rm CR}(R,z) \simeq - \pi^2 \frac{e^2v_{\rm A}}{m_pc^2\Omega_p} RD\frac{\partial f^{\Sigma}}{\partial z} = \nu_{in}(z)\,.
\end{equation}
Hence, we can straightforwardly apply the results of Sec.~\ref{sec:sol_nolos}: by replacing $pf_0(p)$ with $Rf_0^\Sigma(R)$ in Eq.~(\ref{eq:n1}), we immediately derive $n_1(R)$ and $\mathcal{K}(R)$. Deriving $n_2(R)$ is, however, a more complicated task, since $u^{(\alpha)}$ is different for different species, and the boundary condition (\ref{eq:z2_boundary}) cannot be directly applied for $f^{\Sigma}(R)$.

In Appendix~\ref{app1} we obtain an analytical solution for $f^{(\alpha)}(R,z)$ within the diffusion zone, and Eq.~(\ref{eq:S0_general_nuclei}) gives the resulting spectra $f_2^{(\alpha)}(R)$ at the inner boundary. Substituting this in the boundary condition $S^{(\alpha)}_0 = u^{(\alpha)}f_2^{(\alpha)}$ yields the total flux for each species,
\begin{equation}
S^{(\alpha)}_0(R) = \frac{v_{\rm A1}f^{(\alpha)}_0e^{\frac{\mathcal{K}}{3}(\chi-1)}}{\chi^{4/3}\frac{v_{\rm A1}}{u^{(\alpha)}}+
\frac{\mathcal{K}-1}{\mathcal{K}}e^{\frac{\mathcal{K}}{3}(\chi-1)} - \chi + \frac{1}{\mathcal{K}} } \,,
\label{eq:S0_nuclei_final}
\end{equation}
where $\chi(R) = (n_2/n_1)^{-3/2}\leq1$. Combining it with Eq.~(\ref{eq:Snoloss_solution}) leads to an algebraic equation for $\chi$,
\begin{equation}
    \sum Z^{(\alpha)}S^{(\alpha)}_0 = \mathcal{K}S_{\rm D1}\,,
\end{equation}
which finally gives us the value of $n_2(R)$. Then, Eq.~(\ref{eq:S0_nuclei_final}) can be used to calculate fluxes of individual species.

\section{Contribution of heavier nuclei}\label{sec:nucl_contribution}

Knowing the total fluxes of individual species allows us to estimate their contribution to the total excitation rate, Eq.~(\ref{eq:value_of_gamma_nucl}). Consider the outer boundary $n_1(R)$ of the diffusion zone. Each contribution is proportional to the diffusion flux, viz.,
\begin{equation}
\gamma_{\rm CR1}^{(\alpha)}(R) \propto Z^{(\alpha)}(S^{(\alpha)}_0 - v_{\rm A1} f^{(\alpha)}_0)R \,.
\end{equation}
If we introduce an average flux velocity $u^\Sigma$, defined as
\begin{equation}
    u^\Sigma(R) =  \frac{S^\Sigma_0}{f^\Sigma_2} \,,
\end{equation}
then Eq.~(\ref{eq:n2}) can be used to express $\chi - \frac{1}{\mathcal{K}}$ through $u^\Sigma$, 
\begin{equation}
    \chi - \frac{1}{\mathcal{K}} \equiv \frac{v_{\rm A1}}{u^\Sigma} \chi^{4/3} \,,
    \label{eq:chi_expression2}
\end{equation}
and the individual fluxes in Eq.~(\ref{eq:S0_nuclei_final}) can be rewritten as
\begin{equation}
    S^{(\alpha)}_0(R) = v_{\rm A1} f^{(\alpha)}_0 \left[
    \chi^{4/3}\frac{v_{\rm A1}}{u^{(\alpha)}}\left(1 - \frac{u^{(\alpha)}}{u^\Sigma}\right)e^{\frac{\mathcal{K}}{3}(1-\chi)} +
    \frac{\mathcal{K}-1}{\mathcal{K}}
    \right]^{-1} \,.
\label{eq:S0_species}
\end{equation}

\begin{figure}\centering
\includegraphics[width=\columnwidth]{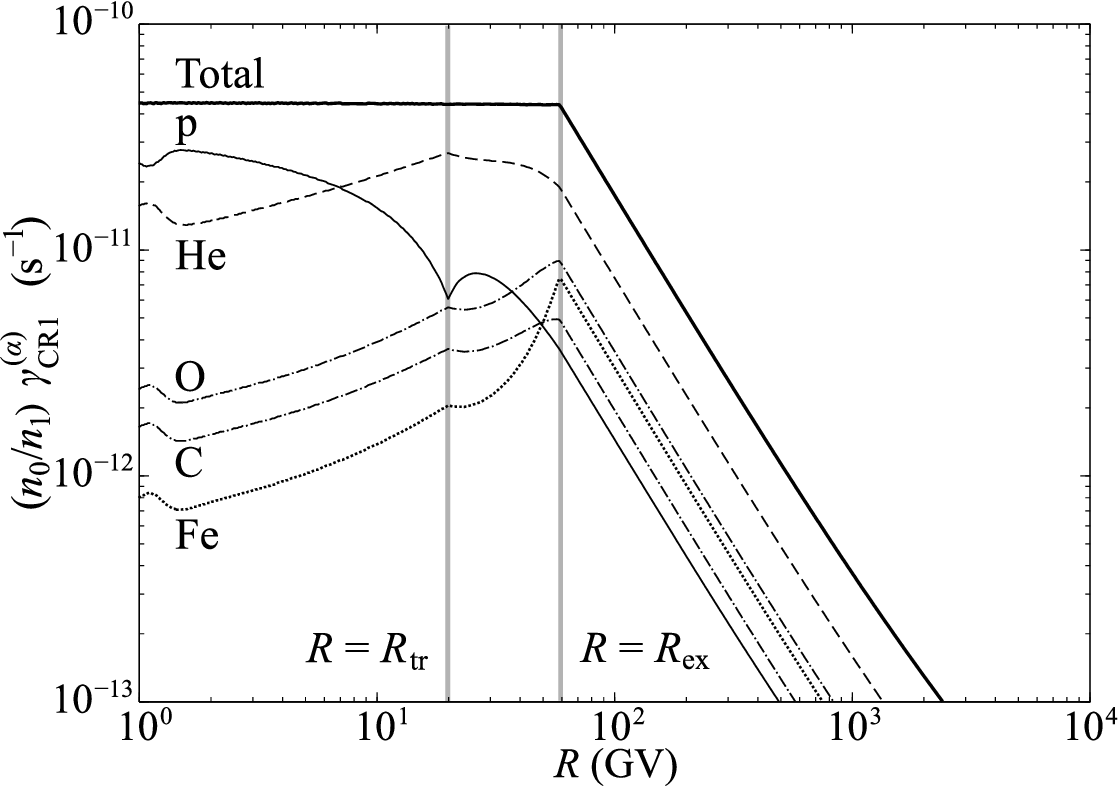}
\caption{Contributions $\gamma_{\rm CR1}^{(\alpha)}$ of different CR nuclei to the total wave excitation rate at the outer boundary of the diffusion zone, plotted versus rigidity $R$. 
The left vertical line indicates the transition rigidity $R_{\rm tr}$, where the outer boundary changes from $n_1=n_{\rm cr}(R)$ to $n_1=n_0$, the right vertical line shows the excitation threshold $R_{\rm ex}$ (see Sec.~\ref{sec:setup} and Fig.~\ref{fig:diagram}).} 
\label{fig:excitation}
\end{figure}

Contributions of different nuclei to the excitation rate as well as the total excitation rate at $n=n_1(R)$ are plotted in Fig.~\ref{fig:excitation} for\footnote{This value of $n_0$ is set smaller than that in Fig.~\ref{fig:gammarays} for illustrative purposes, to widen the gap between $R_{\rm tr}$ and $R_{\rm ex}$ in Fig.~\ref{fig:excitation}.} $n_0 = 0.5$~cm$^{-3}$, assuming $\mathcal{N}=6\times 10^{22}$ cm$^{-2}$ for the gas column density and using Eq.~(\ref{f_ISM}) for the ISM spectra (see Sec.~\ref{sec:gammarays}). The results are multiplied by $n_0/n_1$ in order to highlight the excitation-damping balance $\gamma_{\rm CR1} = \nu_0 (n_1/n_0)$. In the range of $R>R_{\rm ex}$ (bound by the right vertical line), where the excitation-damping balance is no longer satisfied, we plot $\gamma_{\rm CR1}^{(\alpha)}(R) \propto Z^{(\alpha)}(u^{(\alpha)} - v_{\rm A0})Rf^{(\alpha)}_0$.

Let us analyze the partial contributions to the excitation rate. For the chosen values of $n_0$ and $\mathcal{N}$, regime (i) is realized in Fig.~\ref{fig:excitation} with the value of $R_{\rm tr}$ a few times smaller than $R_{\rm ex}$. As long as $R$ is sufficiently small, so that $n_1(R)=n_{\rm cr}(R)$ is substantially larger than $n_0$, we have $u^{(\alpha)} \gg v_{\rm A1}$. In this case $\mathcal{K} = 4$ and, according to Eq.~(\ref{eq:chi_expression2}), $\chi \approx \mathcal{K}^{-1}$. Then the term $\propto\chi^{4/3}$ in Eq.~(\ref{eq:S0_species}) can be dropped, and the partial fluxes reduce to 
\begin{equation}
     S_0^{(\alpha)}(R) \approx \frac43v_{\rm A1} f_0^{(\alpha)} \,.
\label{eq:S0_passive}
\end{equation}
Thus, the contributions of different nuclei depicted in Fig.~\ref{fig:excitation} at low $R$ are simply proportional to their charge density.

As $v_{\rm A1}/u^{(\alpha)}$ increases with $R$, the term $\propto\chi^{4/3}$ in Eq.~(\ref{eq:S0_species}) eventually becomes important, and its effect then depends on the sign of $1 - u^{(\alpha)}/u^\Sigma$. We note that the value of $u^{(\alpha)}$ for protons is typically substantially smaller than that for heavier CR species, since protons do not experience fragmentation. Therefore, $u^{(\alpha)} < u^\Sigma$ is assumed for protons and $u^{(\alpha)} > u^\Sigma$ for heavier nuclei, i.e., the proton contribution to the excitation decreases with $R$ while the contribution heavier species increases, and eventually the excitation is dominated by helium nuclei.\footnote{In general, the proton flux may even become sub-Alvenic, $S^{(p)}_0 < v_{\rm A1}f_0^{(p)}$, making their contribution to the excitation negative.} This behavior is seen in Fig.~\ref{fig:excitation} for the range of $R<R_{\rm tr}$ (bound by the left vertical line). The curves for He, O, C and Fe are similar in this case, since Eq.~(\ref{eq:S0_passive}) remain applicable for all species apart from protons.

For $R> R_{\rm tr}$, we have $n_1(R) = n_0$ and hence $v_{\rm A1}=v_{\rm A0}$. The dependencies $S^{(\alpha)}_0(R)$ are then controlled by $\mathcal{K}(R)$ and $\chi(R)$, showing non-monotonic behavior; specifically, the dependencies are determined by interplay of $\chi^{4/3}e^{\frac{\mathcal{K}}{3}(1-\chi)}$ and $\frac{\mathcal{K} - 1}{\mathcal{K}}$ in Eq.~(\ref{eq:S0_species}). In particular, the curve for protons in Fig.~\ref{fig:excitation} exhibits a local maximum, while for helium it decreases monotonically with $R$. 

\section{Gamma-ray emission from molecular clouds}\label{sec:gammarays}

Recent Fermi LAT observations of nearby giant molecular clouds show deficits in the gamma-ray residual map when the expected diffuse emission is modelled assuming uniformly distributed CRs \citep{yang2023}. The authors pointed out that the observed emission ``holes'' reflect the lack of penetration of $\lesssim10$~GeV CRs into denser regions, and proposed that the CR deficit is caused by slower CR diffusion in the clouds.

We apply the presented model of CR self-modulation to the observations by \citet{yang2023}. The ISM spectra of CR protons and heavier nuclei are approximated by
\begin{equation}
    f^{(\alpha)}_0(R) =\varphi^{(\alpha)}_0\tilde{R}^{a}\left(\frac{\tilde{R}^{-1.55}}{1 + (\tilde{R}/0.7)^{1.3}} +
    \frac{5.3\times10^{-4}\tilde{R}^{-1.8}}{1 + (\tilde{R}/1.3\times10^4)^{1.85}} \right) \,,
    \label{f_ISM}
\end{equation}
where $\tilde{R} = R/(1~{\rm GV})$. The functional dependence of the fit and the abundance factors $\varphi_0^{(\alpha)}$ are chosen to reproduce observational data in the range of $1$~GV~$\lesssim R \lesssim 10^4$~GV \citep{Aguilar21}. The exponent $a$ takes into account spectral hardening for heavier nuclei, so that $a = 0$ for protons and $a = 0.1$ for other nuclei. 

\begin{figure}[ht]\centering
\includegraphics[width=0.45\textwidth]{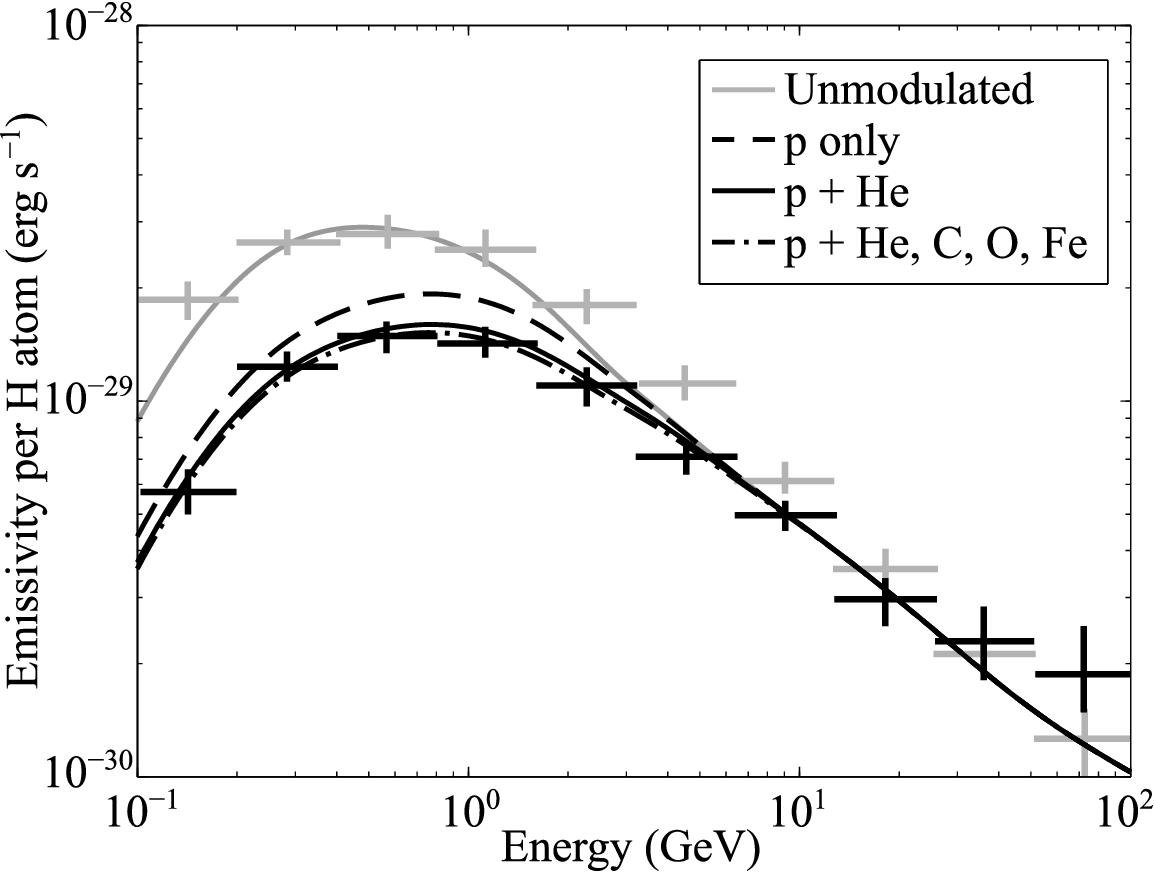}
\caption{Comparison between gamma-ray observations of giant molecular clouds reported by \citet{yang2023} and our theoretical model (see text for parameters). The grey and black symbols represent the measured emissivity derived for the surrounding diffuse gas and dense molecular clumps, respectively. The grey solid line represents the theoretical emissivity computed for the unmodulated ISM spectrum of CRs, Eq.~(\ref{f_ISM}). The black lines show the modulated spectra obtained from our model, with different lines illustrating contributions of different CR nuclei, as indicated in the legend.} 
\label{fig:gammarays}
\end{figure}

We assume $B = 3~\mu$G for the magnetic field in diffuse atomic gas \citep{Jansson2012} and set $n_0 = 1$~cm$^{-3}$ for the border density. The average value of the column density for five dense molecular clumps reported in \citet{yang2023} are estimated as\footnote{A factor of $1.37$ used by \citet{yang2023} to correct for heavy elements is omitted here, since we are interested in the total number of nucleons along the line of sight.} $\mathcal{N} = (m/m_p)/\pi r_{\rm mean}^2\,$, where $m$ and $r_{\rm mean}$ are, respectively, the mass and the mean radius of the clumps listed in their table~1. The resulting values of $\mathcal{N}$ for  clumps $\rm C1$ to $\rm C5$ from that table are, respectively $\approx 1\times 10^{23}$ cm$^{-2}$, $\approx5\times 10^{22}$ cm$^{-2}$, $\approx7\times 10^{22}$ cm$^{-2}$, $\approx5\times 10^{22}$ cm$^{-2}$, and $\approx1\times 10^{23}$ cm$^{-2}$. For our calculations, we employ a conservative value of $\mathcal{N} = 6\times 10^{22}$~cm$^{-2}$. As was shown in Ref.~\citep{dogiel2021}, the net flux velocity $u$ of relativistic CR nuclei penetrating dense molecular clouds is
\begin{equation} 
    u(R,\mathcal{N}) \approx \frac{1}{2}\sigma_{\rm loss}(R) \mathcal{N}c \,,
    \label{eq:uN_definiton}
\end{equation}
where $\sigma_{\rm loss}$ is the cross section of catastrophic losses in the clumps (pion production for protons and fragmentation for heavier nuclei), taken from the GALPROP code (see, e.g., Ref.~\citep{strong1998} and refernces therein).

For the above parameters, we derive the spectra of self-modulated CRs and compute the expected gamma-ray emissivity using parameterization by \citet{Kafex2014}. The results include contributions of CR protons, helium, carbon, oxygen, and iron. To explore the impact of different nuclei, we analyze the following cases: waves are excited by protons only, by protons and helium, and by all included species. 

The results are shown in Fig.~\ref{fig:gammarays} together with the observational data by \citet{yang2023}. 
The computed emissivity is strongly suppressed below $\approx2$~GeV, where helium nuclei significantly affect the results while the contribution of heavier CR species is practically negligible. We see that our results are in excellent quantitative agreement with the observations.

\subsection{Dependence on $\mathcal{N}$, $n_0$, and $B$}

\begin{figure}\centering
\includegraphics[width=0.45\textwidth]{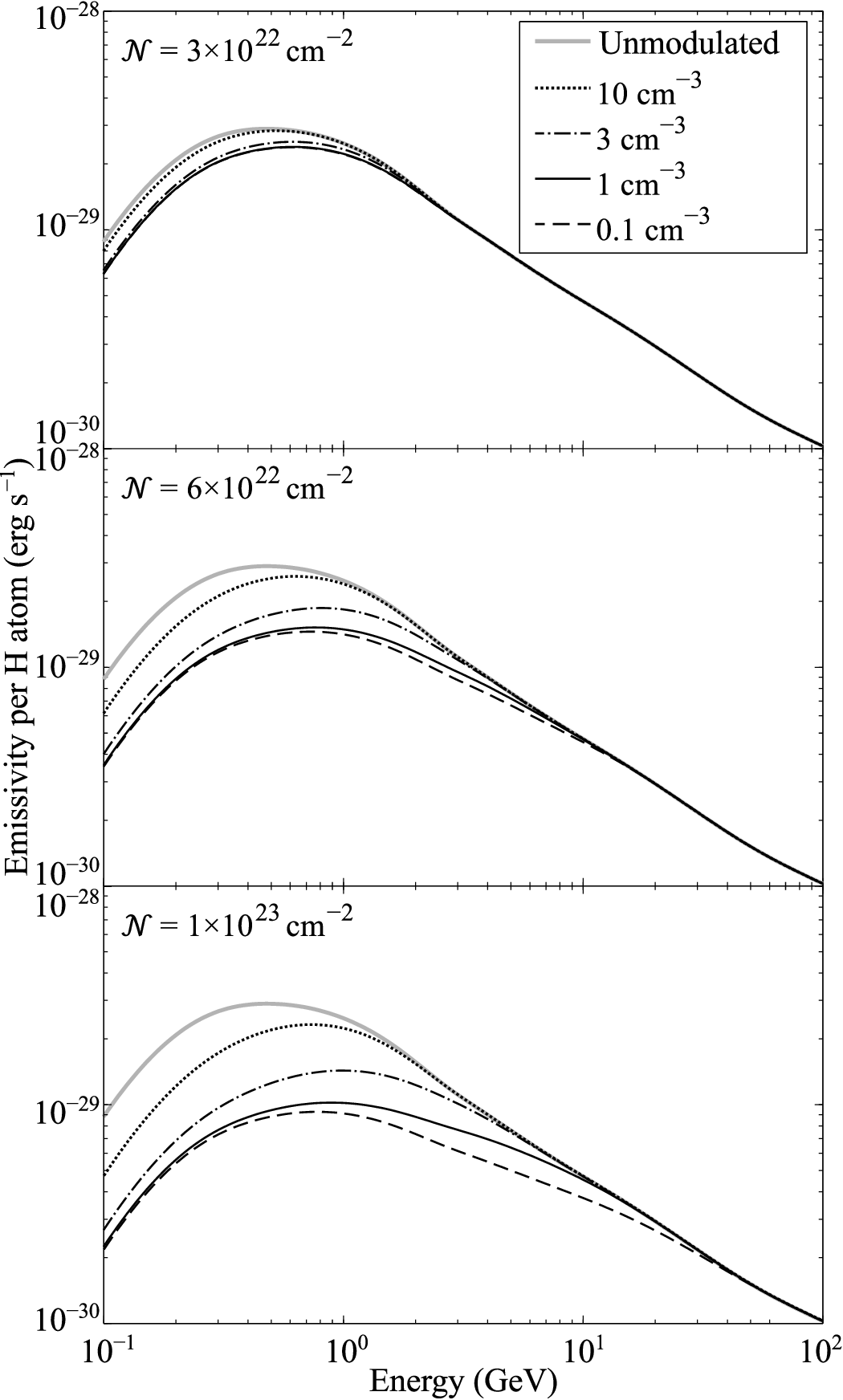}
\caption{Effect of the border density $n_0$ and the cloud column density $\mathcal{N}$ on the gamma-ray emissivity. The interstellar CR spectrum and the magnetic field are the same as in Fig.~\ref{fig:gammarays}, the wave excitation is due to protons and helium nuclei. The three panels show results for three characteristic values of $\mathcal{N}$, the values of $n_0$ are indicated in the legend. The grey solid line shows the emissivity for the unmodulated CR spectrum.} 
\label{fig:flux_gamma_spectra_n0}
\end{figure}

To understand how the column density $\mathcal{N}$ and border density $n_0$ affect the gamma-ray emission, we compute the emissivity for several characteristic values of $\mathcal{N}$ and $n_0$. The results are summarized in Fig.~\ref{fig:flux_gamma_spectra_n0}. 

For $\mathcal{N}=3\times10^{22}$~cm$^{-2}$, regime (ii) of the diffusion zone is realized at lower $n_0$ and regime (i) -- at higher $n_0$ [see Eq.~(\ref{eq:n_trans}) and Fig.~\ref{fig:diagram}]. The CR self-modulation has a relatively small impact on the gamma-ray emission in this case, simply because the excitation threshold $R_{\rm ex}$ is too small and close to the threshold of pion production. 

The impact becomes significant for $\mathcal{N}=6\times10^{22}$~cm$^{-2}$ (corresponds to Fig.~\ref{fig:gammarays}), where regime (i) is primarily realized. We observe a fairly sharp change with $n_0$ occurring around $\sim3$~cm$^{-3}$: the modulation effect tends to saturate at $n_0\leq1$~cm$^{-3}$, whereas for $n_0\geq10$~cm$^{-3}$ it practically vanishes. This is because for larger $n_0$, corresponding to regime (i), the damping rate in the excitation-damping balance is proportional to $n_0$. Since $R_{\rm ex}$ decreases with $n_0$ in this regime, it eventually falls below the range of $R$ contributing to the emission and the modulation effect vanishes. On the other hand, a transition to regime (ii) occurs near the smallest $n_0$, and $R_{\rm ex}$ is then set by intersection of $n_1(R)$ and $n_2(R)$ -- so that the modulated spectrum is no longer dependent on $n_0$.

And for $\mathcal{N}=1\times10^{23}$~cm$^{-2}$, we observe a dramatic suppression which does not saturate as $n_0$ decreases, with the emission reduced by a factor of a few at GeV energies. Now $R_{\rm ex}$ is so high that a very broad range of modulated CR spectrum contributes to the emission, which explains the strength of the effect. At the same time, the diffusion zone is now well in regime (i), and therefore the effect does not saturate within the selected range of $n_0$.    

We note that the reduction in the modulation effect seen with increasing $n_0$ can be on average compensated by increasing $\mathcal{N}$. To quantify this trend, let us consider regime (i) for sufficiently large $\mathcal{N}$ and $n_0$: according to Eq.~(\ref{eq:Snoloss_solution}), the modulated CR flux is then given by $S_0= S_{\rm D0}\propto n_0^{3/2}$ for a broad range of $R_{\rm tr}\leq R\leq R_{\rm ex}$, and the modulated CR spectrum, $S_0/u$ with $u\propto \mathcal{N}$, scales as $n_0^{3/2}/\mathcal{N}$. Eq.~(\ref{p_ex1}) shows that $R_{\rm ex}$ in this regime is also (approximately) a function of $n_0^{3/2}/\mathcal{N}$. Therefore, the resulting gamma-ray emission (which is an integral quantity of the CR spectrum) is approximately similar for similar values of $n_0^{3/2}/\mathcal{N}$.

In the same way, we can assess the effect of the magnetic field strength. For sufficiently large $\mathcal{N}$ and $n_0$, the modulated CR spectrum is independent of $B$ within a broad range of $R_{\rm tr}\leq R\leq R_{\rm ex}$, where also $R_{\rm ex}$ does not depend (approximately) on $B$, and therefore the emission is practically independent of $B$, too. According to Eqs.~(\ref{p_tr}) and (\ref{p_ex1}), decreasing $n_0$ and/or $\mathcal{N}$ (and/or increasing $B$) reduces the gap between $R_{\rm ex}$ and $R_{\rm tr}$ (which is a function of $B$), and thus dependence on $B$ gradually increases. However, the dependence remains rather weak: e.g., by changing $B$ in Fig.~\ref{fig:gammarays} from $3~\mu$G to $5~\mu$G \citep{Jansson2012}, we obtain an equally good fit to the data for $\mathcal{N}\approx8\times10^{22}$~cm$^{-2}$. 

\section{Conclusions}
\label{conclusions}

In this paper we have investigated penetration of relativistic interstellar CRs into molecular clouds surrounded by nonuniform diffuse envelopes. The present work extends and generalizes our earlier model of CR self-modulation \citep{ivlev18, dogiel2018}, where we assumed gas density to be constant across the whole envelope. While the overall qualitative picture remains largely unchanged, several reported findings may have a profound impact on our understanding of the process: 
\begin{enumerate}
    \item Gas density in the envelopes typically increases monotonically going from the ISM to dense molecular clumps, which makes it possible to use the density as the relevant coordinate for the problem of CR penetration. This enables us to obtain a self-consistent {\it analytical} solution of the problem, without assuming the value of density at which the wave excitation by CRs occurs. The resulting solution is {\it universal}, as it does not depend on a particular shape of the spatial density distribution.
    \item The border density $n_0$, through which interstellar CRs enter the envelope, corresponds to a local minimum of the wave damping rate due to ion-neutral collisions. The minimum is associated with a transition between different dominant ions in the envelope and the ISM, so that waves are only exited at $n\geq n_0$. 
    \item The diffusion zone, where CRs are scattered at the self-excited waves, forms as a result of intrinsic inhomogeneity which controls locations of the zone boundaries for a given CR energy. The computed density values for the boundaries turn out to be substantially smaller than those assumed earlier in the constant-density model. This implies a proportionally smaller wave damping and, therefore, more efficient self-modulation for otherwise the same conditions -- thus affecting CRs at substantially higher energies.
    \item The diffusion zone is relatively narrow: the ratio of densities at its inner and outer boundaries is shown not to exceed the value of $2^{4/3}\approx 2.5$. At the same time, the mean free path of CRs due to their scattering at the self-generated turbulence is always much smaller than the inhomogeneity scale and, therefore, the diffusion approximation is always applicable.
    \item The magnitude of flux of self-modulated CRs is always set by the universal diffusion flux $S_{\rm D}$, Eq.~(\ref{eq:sdd_definition}), while in case of uniform envelopes the diffusion component was shown to dominate the flux at higher energies and the avection component -- at lower energies \citep{ivlev18}. Now, we rigorously show that the flux of CRs into the cloud interior never exceeds the value of $4S_{\rm D}$ computed at the outer boundary of the diffusion zone.  
    \item The present model takes into account contributions of different CR nuclei, thus extending the results obtained for constant-density envelopes by \citet{dogiel2018}. Now we show that the relative contribution of helium nuclei to the wave excitation increases with energy and can eventually become dominant.
 \end{enumerate}

We estimated the effect of CR self-modulation on the gamma-ray emission, and showed that the emission can be reduced dramatically at energies below several GeV. The magnitude of this effect is determined by the cloud column density $\mathcal{N}$ as well as by the border density $n_0$ and the magnetic field strength $B$. 

The dependence on $\mathcal{N}$ exhibits a sharp threshold behaviour. For the local Galactic spectrum of relativistic CRs and $B=3~\mu$G, the effect is weak for $\mathcal{N} \leq 3\times 10^{22}$~cm$^{-2}$, while for $\mathcal{N} = 6\times 10^{22}$~cm$^{-2}$ and $1\times 10^{23}$~cm$^{-2}$ the emission at $\lesssim1$~GeV is reduced by a factor of $\approx2$ and $\approx4$, respectively.

The dependence on $n_0$ is determined by the value of $\mathcal{N}$: weakening of the modulation effect with increasing $n_0$ can be on average compensated by increasing $\mathcal{N}$. For sufficiently large $\mathcal{N}$, the gamma-ray emission at $\lesssim1$~GeV is approximately similar for similar values of $n_0^{3/2}/\mathcal{N}$. In the same way, the effect of increasing $B$ is compensated by a comparable increase in $\mathcal{N}$.


We applied our model to explain recent Fermi LAT observations of nearby giant molecular clouds \citep{yang2023}, showing deficits in the gamma-ray emission at GeV energies. For a fairly conservative set of parameters, with the magnetic field of 3--5~$\mu$G and the cloud column density of (6--8)$\times 10^{22}$~cm$^{-2}$, the computed gamma-ray spectra demonstrate an excellent agreement with the observations.


\begin{appendix}

\section{Analytical solution for individual nuclei in the diffusion zone}
\label{app1}

We write the solution of Eq.~(\ref{eq:main_propagation_nucl}) in the following form:
\begin{equation}
    f^{(\alpha)}(R,z) = e^\eta\left(f^{(\alpha)}_0 - S^{(\alpha)}_0 \int\limits_{0}^\eta\frac{e^{-\eta'}}{v_{\rm A} }d\eta'\right) \,,
\label{eq:f_integral_form}
\end{equation}
where
\begin{equation}
    \eta(R,z) = \int\limits_{z_1}^z\frac{v_{\rm A}}{D} dz' \,.
    \label{eq:eta_def}
\end{equation}
Substituting $v_{\rm A}/D$ from Eq.~(\ref{eq:D_noloss}) into Eq.~(\ref{eq:eta_def}) yields 
\begin{equation}
    \eta(R,n) = 2\ln \frac{n}{n_1} - \frac{\mathcal{K}}{3}\left[1 - \left(\frac{n}{n_1}\right)^{-3/2}\right] \,.
\end{equation}
By inserting $\eta$ into Eq.~(\ref{eq:f_integral_form}), we notice that the resulting integral can be calculated analytically and expressed through a new variable $(n/n_1)^{-3/2}$. This gives us the following spectrum at the inner boundary:
\begin{equation}
    f^{(\alpha)}_2(R) = \chi^{-4/3}\left[\left(f^{(\alpha)}_0 - \frac{\mathcal{K}-1}{\mathcal{K}}\frac{S^{(\alpha)}_0}{v_{\rm A1}}\right)e^{\frac{\mathcal{K}}{3}(\chi-1)} + 
    \left(\chi -\frac{1}{\mathcal{K}}\right)\frac{S^{(\alpha)}_0}{v_{\rm A1}}
    \right]\,,
\label{eq:S0_general_nuclei}
\end{equation}
where $\chi(R) = (n_2/n_1)^{-3/2}$. Similarly to what is derived for CR protons in Sec.~\ref{sec:sol_nolos}, we see that also spectra $f^{(\alpha)}_2(R)$ of other nuclei penetrating the cloud do not depend on the shape of $n(z)$. 

\section{Stability of the stationary solution}
\label{app2}

Assume that the stationary solution, which is described by the CR transport equation (without losses), Eq.~(\ref{eq:main_propagation}), and the wave excitation-damping balance, Eq.~(\ref{Wave_Eq}), is disturbed such that $f(z,t) = \bar{f}(z) + \delta f(z,t)$ and $W(z,t) = \bar{W}(z) + \delta W(z,t)$. Here, $\bar{f}$ and $\bar{W}$ denote the derived stationary solution, and the perturbations are assumed to be small, $\delta f \ll \bar{f}$ and $\delta W \ll \bar{W}$. Employing the non-stationary wave equation, $\partial W/\partial t=2(\gamma_{\rm CR}-\nu_{in})W$, and utilizing Eq.~(\ref{D}) to relate $\bar W$ and $\bar D$, we obtain 
\begin{align}
&\frac{\partial\delta f}{\partial t} = \frac{\partial}{\partial z}\left(
\bar{D} \frac{\partial \delta f}{\partial z} - v_{\rm A} \delta f + \frac{6m_p \Omega_p}{vv_{\rm A}p^3}\bar{D}\nu_{in}\delta W
\right)\,, \label{eq:A2-1} \\
&\frac{\partial\delta W}{\partial t} = - \frac{vv_{\rm A}p^3}{3m_p \Omega_p}\frac{\partial \delta f}{\partial z} - 2\nu_{in} \delta W \,.
\label{eq:A2-2}
\end{align}
For small-scale perturbations, both $v_{\rm A}$ and $\bar{D}$ can be treated as constants. Assuming the perturbations $\propto \exp(-i\omega t + ikz)$, the above equations lead to the following dispersion relation:
\begin{equation}
\omega^2 + (i\bar{D}k^2 + 2i\nu_{in} - v_{\rm A}k)\omega - 2i\nu_{in}v_{\rm A}k = 0 \,.
\label{DR}
\end{equation}
To analyze stability of the obtained polynomial, we apply the Hermite-Biehler theorem (see, e.g., Ref.~\citep{Kozhan2024}): all the zeros of a complex polynomial $\alpha(\omega) + il\beta(\omega)$ lie in the lower half-plane if $l>0$ and zeros of the real polynomials $\alpha(\omega)$ and $\beta(\omega)$ strictly interlace. In our case, $\alpha = \omega^2 - v_{\rm A}k\omega$, $\beta = \omega - \frac{2\nu_{in}}{\bar{D}k^2 + 2\nu_{in}}v_{\rm A}k$, and $l = \bar{D}k^2 + 2\nu_{in}$. Since $l > 0$, the zeros of $\alpha$ and $\beta$ are real, and the zero of $\beta$ is strictly between the zeros of $\alpha$, Eq.~(\ref{DR}) has only stable solutions.

\end{appendix}

\end{document}